\documentclass[twocolumn]{emulateapj09}

\usepackage{graphicx}
\usepackage{epsfig}

\slugcomment{To appear in ApJ}

\shorttitle{Measuring Neutron-Star Masses and Radii}
\shortauthors{Psaltis et al.}

\begin{document}


\title{Prospects for Measuring Neutron-Star Masses
and Radii with X-Ray Pulse Profile Modeling}

\author{Dimitrios Psaltis\altaffilmark{1,2}, Feryal
  \"Ozel\altaffilmark{1,2,3}, and Deepto Chakrabarty\altaffilmark{4}}

\altaffiltext{1}{Astronomy Department,
University of Arizona,
933 N.\ Cherry Ave.,
Tucson, AZ 85721, USA}
\email{dpsaltis, fozel@email.arizona.edu, deepto@mit.edu}

\altaffiltext{2}{Institute for Theory and Computation,
Harvard-Smithsonian Center for Astrophysics, 60 Garden St., Cambridge,
MA 02138, USA}

\altaffiltext{3}{Radcliffe Institute for Advanced Study, Harvard
University, 8 Garden St., Cambridge, MA 02138, USA}

\altaffiltext{4}{Department of Physics and Kavli Institute for
  Astrophysics and Space Research, Massachusetts Institute of
  Technology, Cambridge, MA 02139, USA}

\begin{abstract}
Modeling the amplitudes and shapes of the X-ray pulsations observed from
hot, rotating neutron stars provides a direct method for measuring
neutron-star properties. This technique constitutes an important part
of the science case for the forthcoming {\em NICER\/} and proposed {\em
  LOFT\/} X-ray missions.  In this paper, we determine the number of
distinct observables that can be derived from pulse profile modeling
and show that using only bolometric pulse profiles is insufficient for
breaking the degeneracy between inferred neutron-star radius and
mass. However, we also show that for moderately spinning (300--800 Hz)
neutron stars, analysis of pulse profiles in two different energy
bands provides additional constraints that allow a unique
determination of the neutron-star properties. Using the fractional
amplitudes of the fundamental and the second harmonic of the pulse
profile in addition to the amplitude and phase difference of the
spectral color oscillations, we quantify the signal-to-noise ratio
necessary to achieve a specified measurement precision for neutron
star radius. We find that accumulating $10^6$~counts in a pulse
profile is sufficient to achieve a $\lesssim 5$\% uncertainty in the
neutron star radius, which is the level of accuracy required to
determine the equation of state of neutron-star matter. Finally, we
formally derive the background limits that can be tolerated in the
measurements of the various pulsation amplitudes as a function
of the system parameters.
\end{abstract}


\section{INTRODUCTION}

When the surface emission from a spinning neutron star is not uniform,
a periodic brightness oscillation is produced as the hot and cold
spots spin in and out of the line of sight of a distant observer. Such
brightness variations may be caused by the magnetic field topology on
the stellar surface of a pulsar, by the non-uniform thermonuclear
burning on the surface of an X-ray burster, or by the anisotropic
accretion of matter from a companion star.  The amplitudes and shapes
of the resulting pulsations are determined not only by the brightness
contours on the stellar surface but also by the degree of strong-field
gravitational lensing that photons experience on their paths to the
distant observer (Pechenick et al. 1983; Strohmayer et al. 1997). For
this reason, pulse profile modeling is a powerful method for measuring
neutron-star masses and radii (see Strohmayer 2004). An advantage of
this method is that it does not require a measurement of the distance
to the neutron star.  Two future X-ray missions, NASA's approved {\em
  NICER\/} (Gendreau et al.\ 2012) and ESA's proposed {\em LOFT\/}
(Feroci et al.\ 2012), rely on pulse profile modeling to measure the
masses and radii of neutron stars in two classes of sources that show
surface brightness oscillations in the X-rays. {\em NICER\/} targets
the pulsed surface emission that has been detected from
rotation-powered millisecond pulsars, while {\em LOFT\/} is designed
to measure the pulse profiles of accretion-powered millisecond pulsars
and of thermonuclear bursters.

Earlier attempts to measure neutron-star properties from rotation-powered
(e.g., Pavlov \& Zavlin 1997; Bogdanov et al.\ 2007) and
accretion-powered millisecond pulsars (e.g., Poutanen \& Gierlinski
2003; Leahy et al.\ 2008) and bursters (Nath et al.\ 2002) resulted in
large, correlated uncertainties between the inferred masses and radii.
{\em NICER\/}'s design, which will allow accumulating a large number of counts
for each of its targets over very long integration times, and {\em LOFT}'s
large collecting area, which will lead to highly accurate pulse
profiles even during the course of a 10~s X-ray burst, will address
the issue of reducing the statistical uncertainties of the
measurements. However, even when the statistical errors are reduced,
significant correlations between the inferred parameters remain. This
has been recognized in earlier studies (e.g., Nath et al.\ 2002; Poutanen 
\& Beloborodov 2006) and has been demonstrated more recently in a detailed 
study of parameter estimation using mock {\em LOFT} data (Lo et al.\ 2013).

In this paper, we use simulated pulse profiles from spinning neutron
stars in order to identify the origin of the degeneracies in the
measurements of masses and radii that are obtained with this
technique. We use a Fourier series decomposition of the pulse profiles
at different photon energies to quantify the number of distinct
observables that can be measured from each profile. We show that,
because gravitational lensing suppresses the amplitudes of the high
harmonics, this number is rather small and is practically independent
of the number of phase bins used in the measurement. However, the
number of independent parameters that are required to uniquely
characterize each system is rather large. Therefore, the effective
number of degrees of freedom in comparing theoretical models to data
is very small or zero, causing the observed correlations between
parameters.

We further demonstrate that obtaining pulse profiles at different photon
energies significantly reduces the extent of these correlations. This
is because the modulation of the spectrum due to Doppler effects at
moderate spin frequencies introduces a photon-energy dependent
structure to the pulse profiles. Observing, therefore, pulse profiles
in multiple energy bands leads to measuring additional, uncorrelated
observables, thereby increasing the effective number of degrees of
freedom.  The Fourier series decomposition approach that we present
here can be used in defining the optimal ranges of photon energies and
in formulating analysis strategies that maximize the effective number
of degrees of freedom. Moreover, it provides a useful
order-of-magnitude estimate for the number of photons that are
required to be accumulated and for the level of background that can be
accommodated in order for a specified precision to be reached in radius
measurements.

\section{CALCULATIONS OF PULSE PROFILES}

We use the ray tracing algorithm described in Psaltis \& \"Ozel (2013)
to calculate the brightness oscillations detected by an observer at
infinity that arise from a circular, uniform hot spot of angular
radius $\rho$ on the surface of a spinning neutron star.  The observer
and the center of the spot are located at an inclination $i$ and
at a colatitude $\theta_{\rm s}$, respectively, with respect to the
stellar spin axis.  We assume that the emission from the hot spot has
a blackbody spectrum and is isotropic in the local Lorentz frame on
the stellar surface. Note that, in general, the beaming of the
emerging radiation may not be isotropic, but will depend on the
particular type of system under consideration (see also end of \S3.1).

The targets of interest for missions such as {\em NICER\/} and {\em
  LOFT\/} spin at $\sim 200-700$~Hz.  The spacetime around such
neutron stars can be uniquely described by the Hartle-Thorne metric
(Hartle \& Thorne 1968). Calculations within this metric allow us to
accurately account for the effects of Doppler shifts and aberration,
of frame dragging, as well as of the oblateness of the stellar surface
and of its quadrupole moment\footnote{Simulations of pulse profiles
  for neutron stars spinning at $\gtrsim 700$~Hz can only be performed
  with numerical spacetimes, which depend on the details of the
  equation of state (see Cadeau et al.\ 2007 and discussion in Psaltis
  \& \"Ozel 2013).}. Morsink et al.\ (2007) and Psaltis \& \"Ozel
(2013) showed that all these effects need to be taken into account in
order for measurements of neutron-star masses and radii via pulse
profile modeling to reach the $\sim 5-10$\% accuracy required to
distinguish between equations of state (e.g., Lattimer \& Prakash
2001; \"Ozel \& Psaltis 2009).

In this setup, eight distinct parameters are required to fully specify
the geometry of the system and the spacetime of a neutron star
spinning at a known frequency $f$:\\ 
\hspace*{0.5cm}{\em (i)\/} the mass $M$ of the neutron star;\\ 
\hspace*{0.5cm}{\em (ii)\/} the equatorial radius $R_{\rm eq}$ of the neutron
star;\\ 
\hspace*{0.5cm}{\em (iii)\/} the ellipticity of its surface
$\epsilon_{\rm s}$;\\
\hspace*{0.5cm}{\em (iv)\/} its specific spin angular momentum $a\equiv
2\pi I f c/GM^2$, where $I$ is its moment of inertia;\\ 
\hspace*{0.5cm}{\em (v)\/} the quadrupole moment of its spacetime as measured 
by the parameter $\eta$;\\ 
\hspace*{0.5cm}{\em (vi)\/} the observer inclination $i$;\\
\hspace*{0.5cm}{\em (vii)\/} the colatitude of the spot $\theta_{\rm s}$; and \\ 
\hspace*{0.5cm}{\em (viii)\/} the angular radius of the spot $\rho$.

If the emission originates from two localized hot spots, as in
the case of polar-cap emission from rotation-powered millisecond
pulsars (see, e.g., Bogdanov et al.\ 2007), then up to two additional
angles may be needed to specify the relative position of the two spots
on the stellar surface.

\begin{figure}[t]
\psfig{file=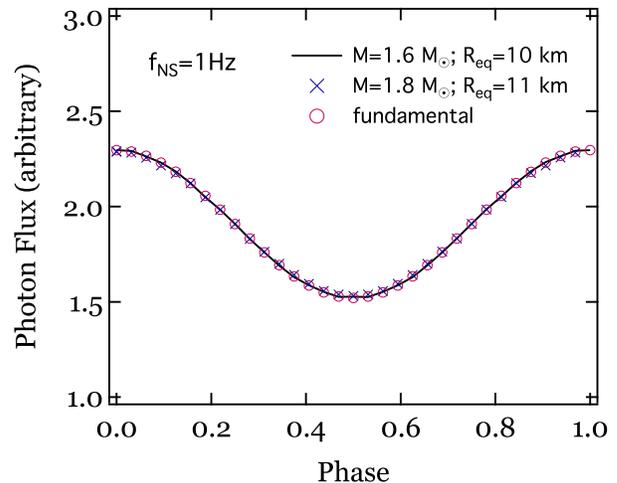,width=3.5in,clip=}
\caption{Pulse profiles (photon flux as a function of rotational
  phase) generated by a circular hot spot on the surface of a neutron
  star spinning at 1~Hz. The solid line corresponds to a neutron star
  with $M=1.6 M_\odot$ and $R_{\rm eq}=10$~km and the crosses to a
  neutron star with $M=1.8 M_\odot$ and $R_{\rm eq}=11$~km, where we
  chose the masses and radii so that the two stars have comparable
  $M/R_{\rm eq}$. The hot spot has a radius of $\rho=10^\circ$, is
  located at a colatitude of $\theta_{\rm s}=40^\circ$, and is
  observed from an inclination of $i=30^\circ$ with respect to the
  stellar spin axis.  In order to demonstrate that, at low spin
  frequencies and for a wide range of geometries, the pulse profile is
  nearly sinusoidal, we use open circles to show the pulse profile for
  the 1.6~$M_\odot$ star, when we have suppressed all the harmonics
  beyond the fundamental. The similarity between all three pulse
  profiles demonstrates visually that pulse-profile modeling for
  slowly spinning neutron stars suffers from a large degeneracy
  between the inferred mass and radius.}
\label{fig:profiles}
\end{figure}

\begin{figure}[t]
\psfig{file=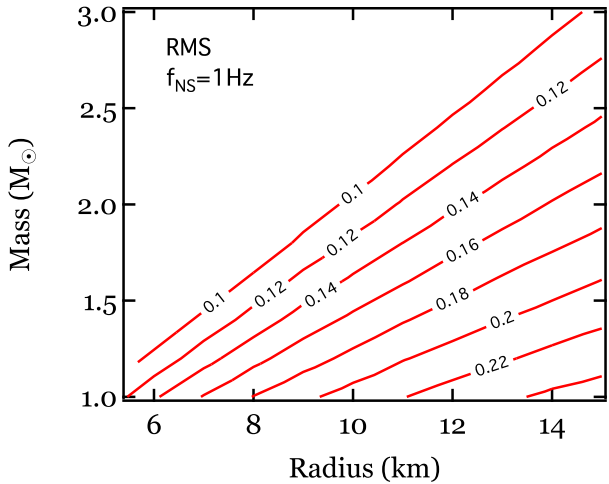,width=3.5in,clip=}
\caption{Contours of constant fractional rms amplitude of pulsations
  generated by a hot spot on the surface of a neutron star spinning at
  1~Hz, as a function of the assumed stellar mass and radius. The
  remaining parameters of the calculation are the same as in
  Figure~\ref{fig:profiles}. The contours lie along lines of constant
  $M/R_{\rm eq}$. This is expected given that the fractional rms
  amplitude depends on the amount of gravitational lensing experienced
  by the photons and the latter, for a slowly spinning neutron star,
  depends only on $M/R_{\rm eq}$.}
\label{fig:rms_1Hz}
\end{figure}

The observed pulse profiles are affected by each of these eight
parameters and could, in principle, contain adequate structure to
allow for uncorrelated measurements of all of them. However, as we
will show in the following section, gravitational light bending smears
the profiles and effectively erases some of the structure that encodes
the detailed properties of the neutron star and of the spacetime. As a
result, realistic pulse profiles do not contain enough information to
measure these eight parameters independently, even at the
signal-to-noise ratios expected when a large number of 
photons is collected.

However, tight relations exist between several of the above
macroscopic quantities that depend very weakly on the equation of
state (e.g., Morsink et al.\ 2007; Yagi \& Yunes 2013; Baub\"ock et
al.\ 2013) and can be used to reduce the number of free parameters
that are necessary to model pulse profiles. In particular, hereafter,
we will use relations that connect the parameters $\epsilon_{\rm s}$,
$a$, and $\eta$ to $M$ and $R_{\rm eq}$ (Baub\"ock et
al.\ 2013). Finally, when the angular size of the spot is small
($\rho\lesssim 10^\circ$), the pulse profile does not depend on this
parameter (see, e.g., Bogdanov et al.\ 2007). Therefore, for systems
in which the surface emission is highly localized, as is expected to
be the case during the first fraction of a second of an X-ray burst
before the burning front has propagated to a significant distance away
from the ignition point (Strohmayer et al.\ 1997, 1998) and for
polar-cap heating in the case of rotation-powered pulsars (e.g.,
Bogdanov 2013), the spot size can be eliminated as a parameter. As a
result, the pulse profile is determined only by four parameters: $M$,
$R_{\rm eq}$, $i$, and $\theta_{\rm s}$.  In the following section, we
show that these four parameters can be independently inferred from
realistic pulse profiles if we use neutron stars that spin at moderate
rates and utilize the photon-energy dependence of the profiles.

\begin{figure}[t]
\psfig{file=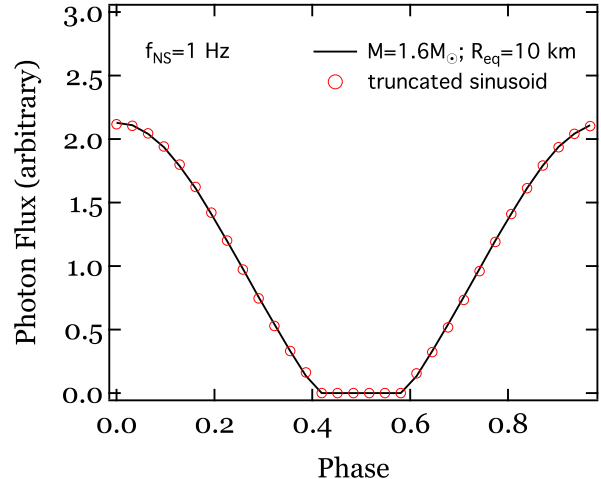,width=3.5in,clip=}
\caption{Pulse profiles generated by a circular hot spot on the
  surface of a neutron star spinning at 1~Hz. In this case, the hot
  spot has a radius $\rho=10^\circ$, is located at a colatitude
  $\theta_{\rm s}=80^\circ$, and is observed from an inclination
  $i=90^\circ$ with 
  respect to the spin axis, such that it is occulted
  by the neutron-star surface for a fraction of the spin period. The
  solid line shows a truncated sinusoid that best describes the result
  of the ray-tracing calculation. Even though the occultation generates
  a large number of measurable harmonics, the pulse profile can be
  accurately described by only two numbers: its amplitude and the
  duration of the occultation.}
\label{fig:profile_occult}
\end{figure}

\section{Measuring Neutron-Star Parameters from Pulse Profile Modeling}

\subsection{Slowly Spinning Neutron Stars}

The external spacetime of a slowly spinning neutron star depends only
on its compactness $GM/Rc^2$. Therefore, modeling pulse profiles
observed from such systems can only lead to a measurement of $M/R_{\rm
  eq}$ and not of the two parameters independently. We illustrate this
degeneracy in Figure~\ref{fig:profiles}, where we show the pulse
profiles from two slowly spinning (1~Hz) neutron stars, with
substantially different masses and radii but with very similar
compactness.  In Figure~\ref{fig:rms_1Hz}, we further demonstrate the
degenerate dependence of pulse profiles on $M/R_{\rm eq}$. In
particular, we plot contours of constant fractional root-mean-squared
(rms) amplitude on the mass-radius plane for a neutron star spinning
at 1~Hz, while keeping fixed the inclination of the observer to
$i=30^\circ$ and the colatitude of the hot spot to $\theta_{\rm
  s}=40^\circ$. As expected, the contours are lines of constant
$M/R_{\rm eq}$.  In the next section, we will discuss how these
contours change as a function of the neutron-star spin frequency.

\begin{figure}[t]
\psfig{file=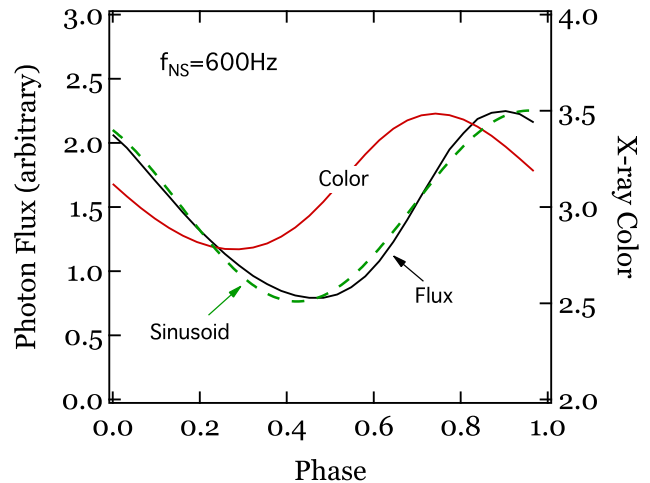,width=3.5in,clip=}
\caption{The pulse profile and the phase dependence of a spectral
  color for a neutron star spinning at 600~Hz. In this calculation,
  the colatitude of the spot is $\theta_{\rm s}=40^\circ$ and the
  inclination of the observer is $i=60^\circ$. The spectral
  color is defined here as the ratio of the number of photons with
  energies above the temperature of the blackbody emission to the
  number of those below. The peak of the spectral color occurs close
  to the phase at which the tangential velocity of the surface is
  maximum. On the other hand, the peak of the radiation flux occurs
  close to the phase at which the projected area of the hot spot is
  maximum. For this reason, the former precedes the latter. The dashed
  line shows the sinusoid that has the same amplitude as that of 
  the fundamental harmonic of the oscillations.}
\label{fig:color}
\end{figure}

\begin{figure}[t]
\psfig{file=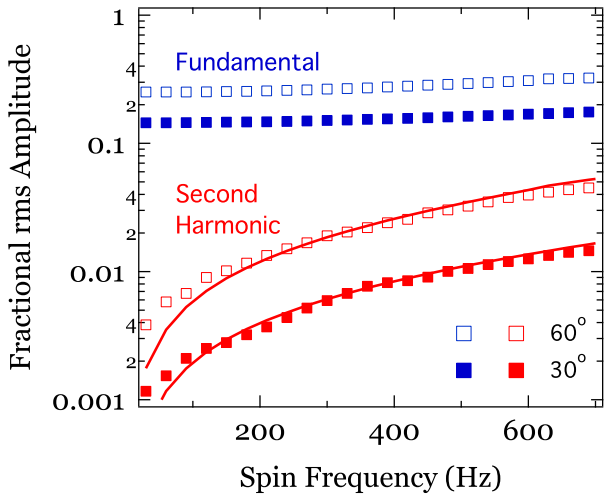,width=3.5in,clip=}
\caption{Amplitudes of the fundamental and second harmonic for the
  pulsations originating on the surface of a 1.6~$M_\odot$, 10~km
  neutron star, as a function of its spin frequency, for two different
  inclinations of the observer (30$^\circ$ and 60$^\circ$). The
  remaining parameters of the 
  calculation are the same as in Figure~\ref{fig:profiles}. At slow
  spins, the pulse profiles are highly sinusoidal whereas, at higher
  spins, Doppler effects introduce asymmetries to the pulse profiles
  and increase the amplitudes of the higher harmonics. Even at spin
  frequencies as high as 600~Hz, the amplitude of the second harmonic
  is $\simeq 10$ times lower than that of the fundamental for these
  geometries. The lines show the approximate scaling of
  equation~(\ref{eq:scaling}) for the amplitude of the harmonic, given
  the calculated amplitude of the fundamental. The amplitude of the
  fundamental increases slightly with spin frequency because of the
  increase in the peak-to-peak excursion caused by Doppler boosts.}
\label{fig:rms_spin}
\end{figure}

Figure~\ref{fig:profiles} also demonstrates that the pulse profile is
highly sinusoidal by comparing the result of the ray-tracing
calculation to a pure sinusoid with the appropriate phase and
amplitude. This implies that, even if the signal to noise of an
observation allows splitting the observed pulse profile into a large
number of phase bins, the complete information content in the profile
is captured by a single quantity: the amplitude of the sinusoid.  
In other words, if we decompose the pulse profile of a slowly spinning
neutron star into a Fourier series, only the amplitude of the
fundamental will be measurable. 

When the geometry is such that the hot spot is occulted by the stellar
surface for a fraction of the spin phase, a large number of harmonics
will be present. However, these additional harmonics acquire large
amplitudes only because of the truncation of the otherwise sinusoidal
profile due to the occultation (Gibbs phenomenon). In this case, the
total information content in the pulse profile is represented by only
two quantities, e.g., the fractional rms amplitude of the oscillation and the
duration of the occultation (see Fig.~\ref{fig:profile_occult}). 

\begin{figure*}[t]
\centerline{
\psfig{file=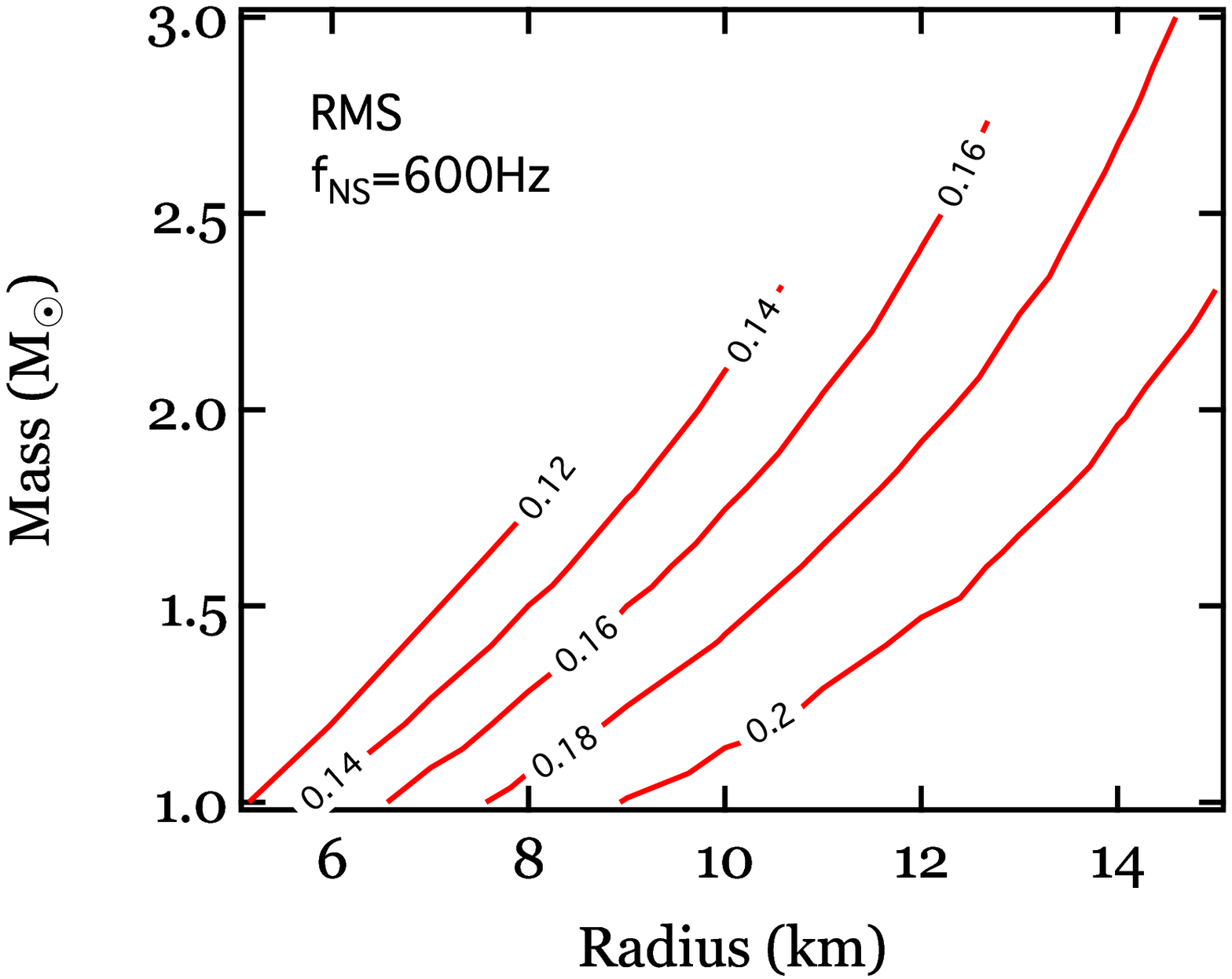,width=3.5in,clip=}
\psfig{file=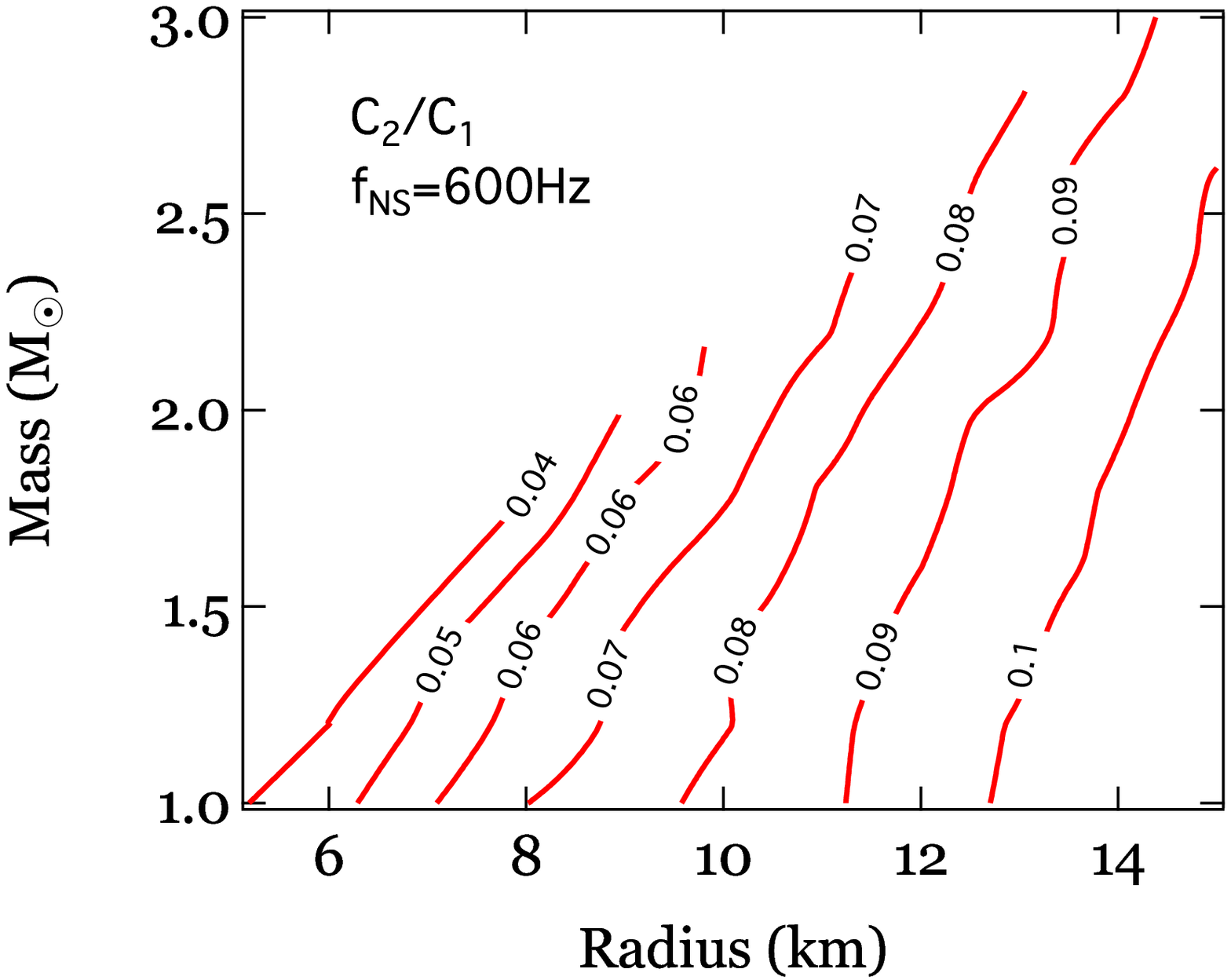,width=3.5in,clip=}}
\caption{Contours of constant {\em (Left)\/} fractional rms amplitude
  and {\em (Right)\/} ratio of the amplitude of the harmonic to that of
  the fundamental for pulsations generated by a hot spot on the
  surface of a neutron star spinning at 600~Hz, as a function of the
  stellar mass and radius. The remaining parameters of the
  calculation are the same as in Figure~\ref{fig:profiles}. For a
  fixed neutron-star spin frequency, the stellar radius determines the
  magnitude of the Doppler effects, which themselves determine
  predominantly the harmonic content of the pulse profiles. For this
  reason, the contours shown in the right panel are nearly vertical
  and correspond to lines of nearly constant radius.}
\label{fig:rms_600Hz}
\end{figure*}

\begin{figure*}[t]
\centerline{
\psfig{file=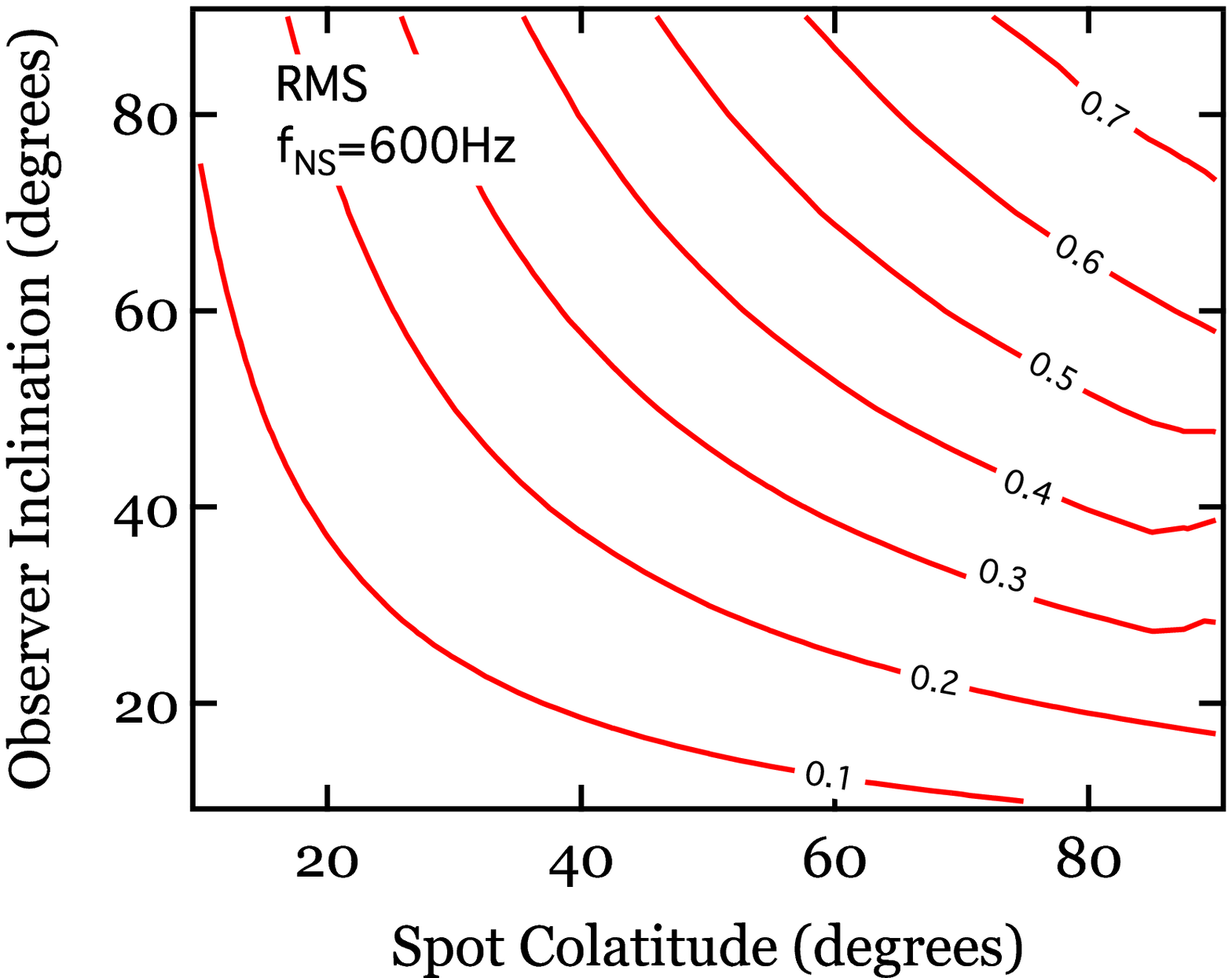,width=3.5in,clip=}
\psfig{file=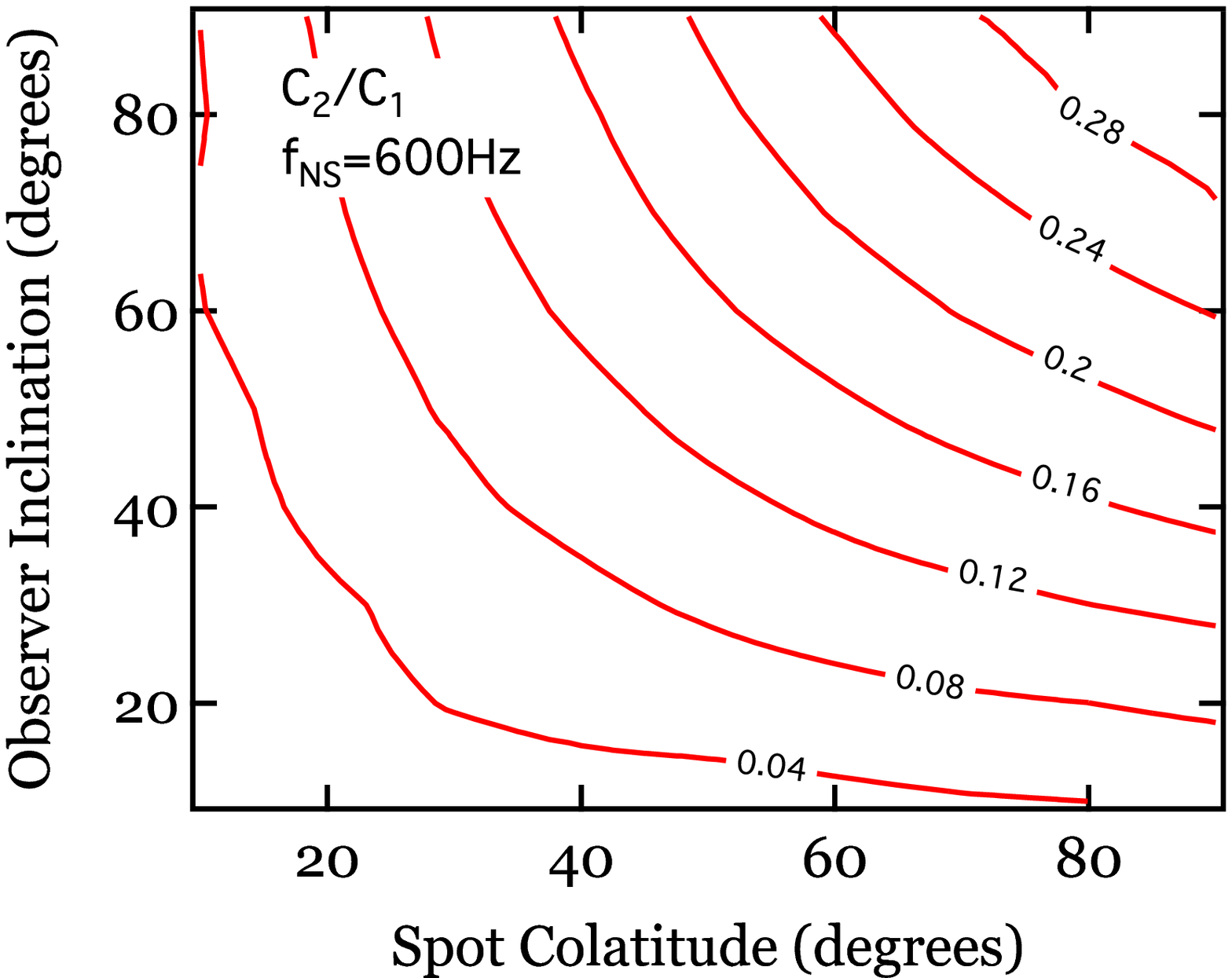,width=3.5in,clip=}}
\caption{Contours of constant {\em (Left)\/} fractional rms amplitude and {\em
    (Right)\/} ratio of the amplitude of the harmonic to that of
  the fundamental for pulsations generated by a hot spot on the
  surface of a neutron star spinning at 600~Hz, as a function of the
  colatitude of the spot and the inclination of the observer. 
  The remaining parameters of the calculation are the same as in 
  Figure~\ref{fig:profiles}. In both cases, the contours lie 
  primarily along curves on which the product $\sin i \sin\theta_{\rm s}$ 
 is constant.} 
\label{fig:rms_600Hz_theta}
\end{figure*}

When we compare the number of unique pieces of information encoded in
the pulse profile of a slowly spinning neutron star (i.e., two in a
geometry with occultation or one without) to the number of parameters
required to describe the system (i.e., three: $M/R_{\rm eq}$,
$i$, and $\theta_{\rm s}$), it becomes apparent that pulse
profile modeling in such a system can only result in highly correlated
measurements of its parameters. Additional structure in the pulse
profiles may also be present due to an anisotropic beaming of
radiation on the stellar surface (e.g., Poutanen \& Beloborodov 2006).
However, measuring the amplitudes of the higher harmonics in this case
will only provide information about the beaming of radiation and not
about the neutron-star properties.

\subsection{Moderately Spinning Neutron Stars}

When a neutron star is spinning at moderate rates ($\sim 300-800$~Hz),
the nearly relativistic velocity of its surface causes three phenomena
that introduce complexity to the pulse profiles: Doppler shifts of the
photon energies, aberration in their angular distribution, and time
delays between photons emitted at different spin phases. We show in
Figure~\ref{fig:color} the deviation of the pulse profile from a pure
sinusoid for a spot at 40$^\circ$ colatitude on a neutron star
spinning at 600~Hz, and observed from a 30$^\circ$ inclination. As found
in earlier studies (e.g., Braje et al.\ 2000), the pulse profile
becomes asymmetric and peaks at an earlier phase compared to the sinusoid.

We quantify the degree of structure in the pulse profiles by comparing
the amplitude of the second harmonic to that of the fundamental as a
function of the spin frequency in Figure~\ref{fig:rms_spin}. As
expected, the amplitude of the second harmonic increases significantly
with increasing spin frequency and is more than an order of magnitude
larger for a 600~Hz star compared to a slowly spinning one. The
harmonic amplitudes also depend strongly on the observer's inclination
and are much larger for an observer located closer to the rotational
equator. Indeed, the ratio of the harmonic amplitude to that of the
fundamental scales approximately as (see Poutanen \& Beloborodov 2006)
\begin{eqnarray}
\frac{C_2}{C_1}&\simeq&2\left(\frac{2\pi f R_{\rm eq}}{c}\right)
\sin i \sin\theta_{\rm s}\nonumber\\
&=&0.126\left(\frac{f}{300~\mbox{Hz}}\right)
\left(\frac{R_{\rm eq}}{10~\mbox{km}}\right)
\sin i \sin\theta_{\rm s}\;.
\label{eq:scaling}
\end{eqnarray}
This approximate scaling is shown in Figure~\ref{fig:rms_spin}, for
two different observer inclinations, and matches the results of the
numerical calculation. Note that the primary scaling is due to the
first-order Doppler effect while higher-order corrections (due to the
oblateness and the quadrupole moment of the neutron star) affect
primarily the numerical factor in this last equation.  Using the
simpler Schwarzschild+Doppler approximation (e.g., Miller \& Lamb
1998; Poutanen \& Beloborodov 2006; Lo et al. 2013), as opposed to the
Hartle-Thorne metric we use here, therefore, leads only to a
systematic bias in the measurement and not to qualitatively different
uncertainties.

It is evident from equation~(\ref{eq:scaling}) that a marked
difference between slowly and moderately spinning neutron stars is
that in the latter case, the amplitude of the second harmonic shows a
strong dependence on the neutron star radius.  The amplitudes of
higher harmonics can, therefore, be useful for breaking the degeneracy
between the stellar mass and radius that we discussed in \S3.1. This
is shown in Figure~\ref{fig:rms_600Hz}, in which contours of constant
fractional rms amplitude (left panel) and of the ratio of the amplitudes of
the second harmonic to the fundamental (right panel) are plotted on
the mass-radius parameter space. In this calculation, the neutron star
is spinning at 600~Hz, the inclination of the observer is
$i=30^\circ$, and the colatitude of the hot spot is
$\theta_{\rm s}=40^\circ$. (Note that this geometry is far from the
one that maximizes the ratio $C_2/C_1$.) The contours of constant
fractional rms amplitude (left panel) are primarily along lines of constant
compactness, as in the case of slowly spinning neutron stars. However,
they bend upwards at large radii because of the increase in the
peak-to-peak flux excursion caused by Doppler boosts. In contrast,
the contours of constant amplitude ratios (right panel) are primarily
vertical, since the harmonic amplitudes increase with stellar radius,
as shown in eq.~[\ref{eq:scaling}]). The weak mass dependence of these
contours arises from the effects of gravitational lensing and from the
redshift factors that need to be taken into account when computing the
velocity of the stellar surface in the local Lorentz frame.

Figure~\ref{fig:rms_spin} demonstrates that, even at the high end of
the observed spin frequencies, the ratio of the amplitude of the
second harmonic to that of the fundamental is quite small; naturally,
the amplitudes of the higher harmonics are even smaller (see, e.g.,
Poutanen \& Beloborodov 2006). Moreover, the right panel of
Figure~\ref{fig:rms_600Hz} shows that the ratio of the amplitude of
the harmonic to that of the fundamental needs to be measured to a
$\sim 10$\% fractional accuracy in order for the observations to
distinguish between neutron-star radii that differ by $\sim 1$~km. As
we will show in \S4, this is quite a severe requirement and makes it
unlikely that future observations will be able to extract more than
two measurable quantities from bolometric pulse
profiles.\footnote{This number can be increased by one if the
  geometry of the system is such that the hot spot is occulted for a
  fraction of the spin phase (see the discussion in \S3.1 for the case
  of slowly spinning neutron stars).}  Therefore, even in this case,
the independent pieces of information in realistic measurements still
falls short of the number of system parameters (i.e., four) that need
to be determined. In principle, the relative phase of the harmonic and
the fundamental provides additional information. However, in practice,
the phase is too poorly determined due to the weakness of the harmonic.

In Figure~\ref{fig:rms_600Hz_theta}, we take a different cut through
the four-dimensional parameter space and plot the dependence of the
fractional rms amplitude (left panel) and the ratio of the amplitude of the
harmonic to that of the fundamental (right panel) as a function of the
spot colatitude and the observer's inclination. In both panels, the
contours of constant amplitude and the contours of constant harmonic
ratio lie along curves on which the product $\sin i
\sin\theta_{\rm s}$ is nearly constant. This implies that, if there is
no occultation, the two measurable quantities cannot be used to infer
the two angles independently. However, given that constraining these
two angles independently is often not of interest, the above
combination $\sin i \sin\theta_s$ can be treated as a single
nuisance parameter, thus reducing the number of system parameters that
need to be measured to three.  We will now show that the photon energy
dependence of the pulse profiles can provide the additional pieces of
information and break the remaining degeneracy between parameters.

\begin{figure}[t]
\psfig{file=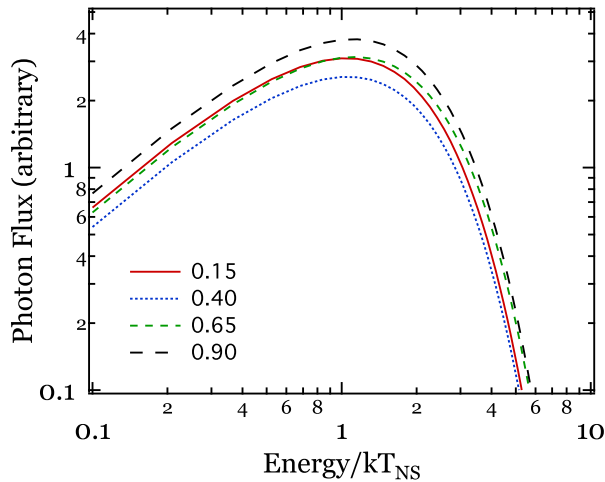,width=3.5in,clip=}
\caption{The evolution of the radiation spectrum, as observed at
  infinity at an inclination of $30^\circ$, generated by a hot spot 
  on the surface of a neutron star spinning at 600~Hz. The other 
  parameters of the calculation are the same as in Figure~\ref{fig:rms_spin}. 
  The various curves correspond to different spin pßhases, with zero 
  representing the phase at which the center of the hot spot and the 
  observer are on the same meridian. Comparing the ordering of the 
  curves at very low and very high photon energies reveals a strong 
  color oscillation during a spin cycle.}
\label{fig:spectrum}
\end{figure}

\begin{figure*}[t]
\centerline{
\psfig{file=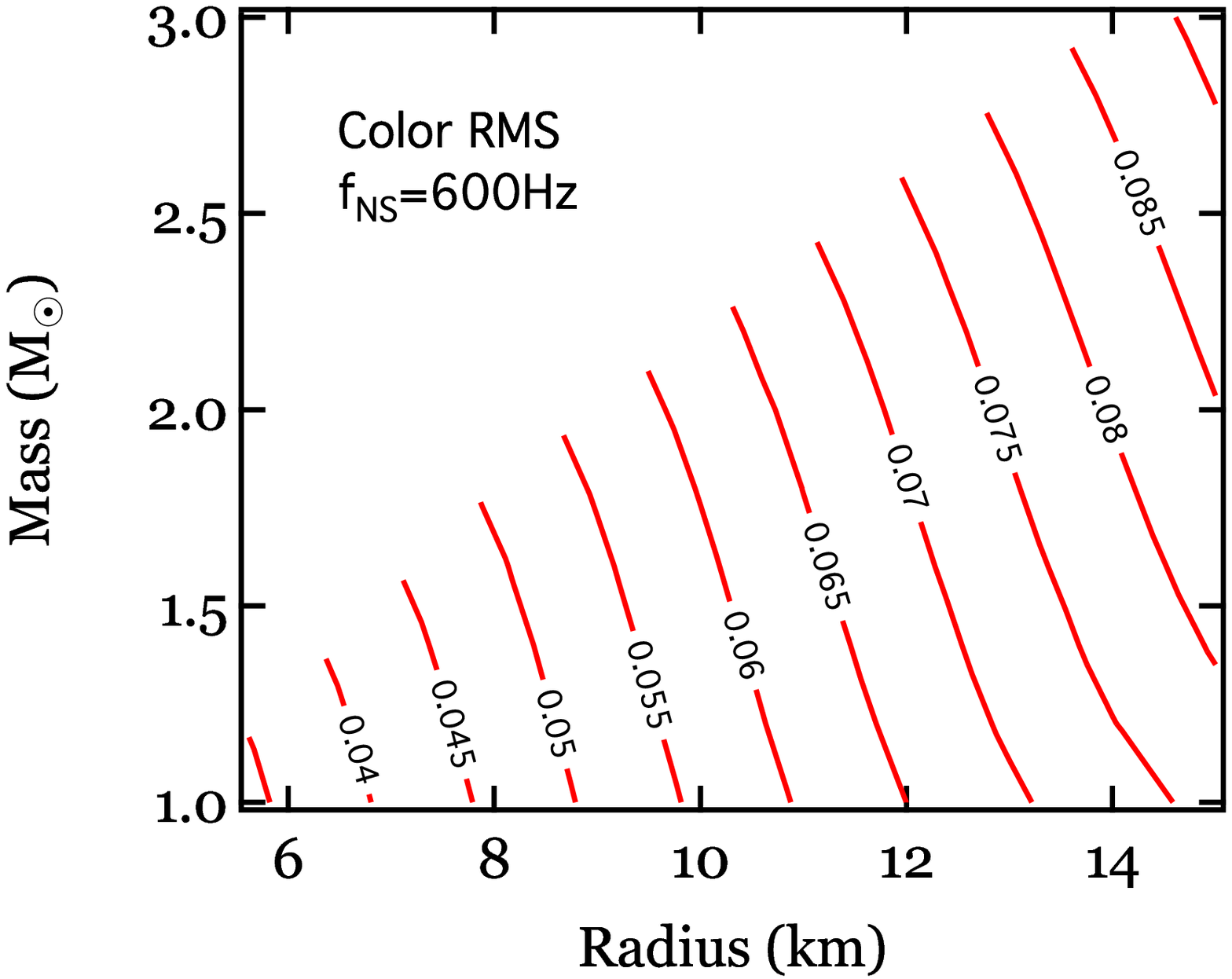,width=3.5in,clip=}
\psfig{file=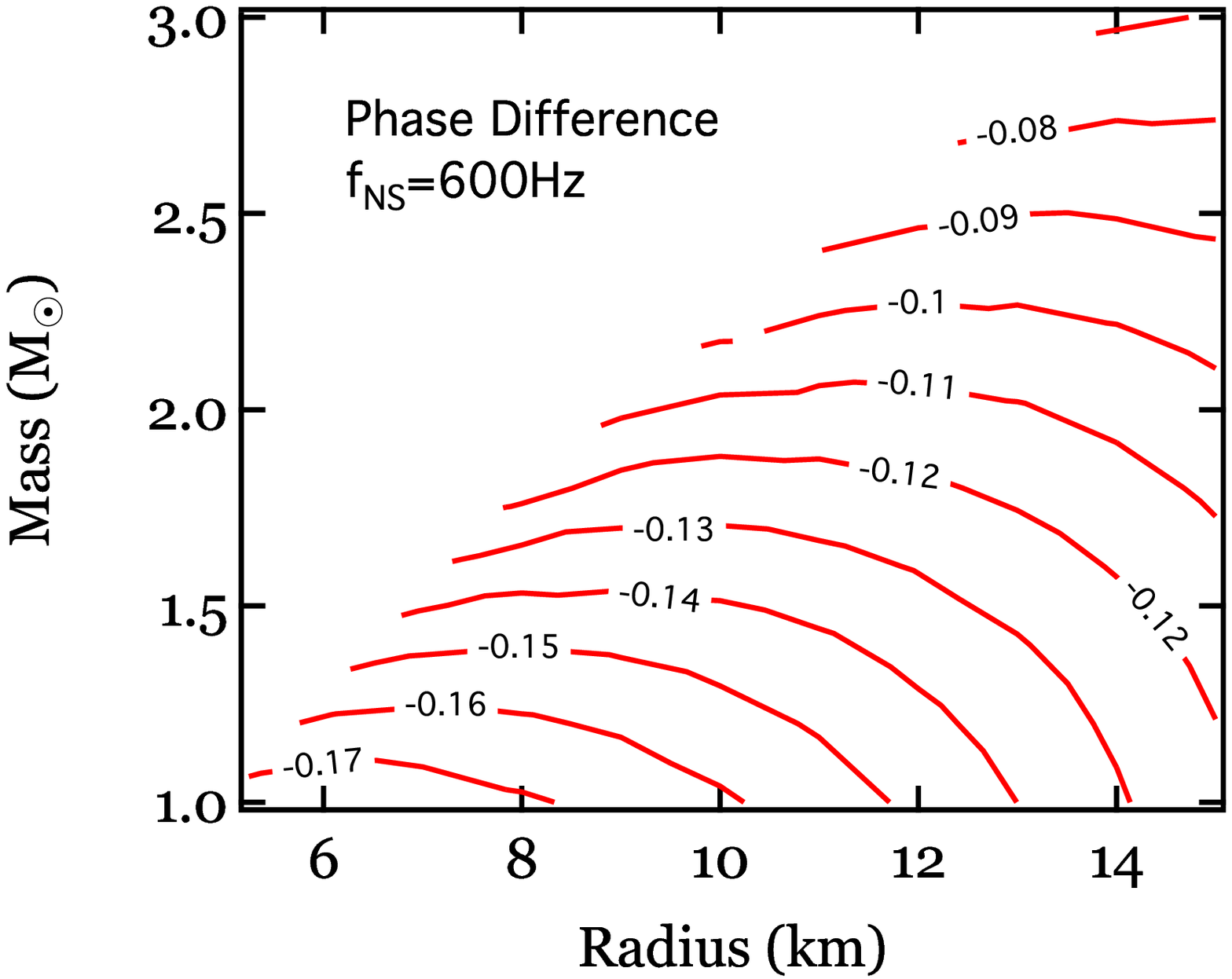,width=3.5in,clip=}}
\caption{Contours of constant {\em (Left)\/} fractional rms amplitude of a
  spectral color oscillation and {\em (Right)\/} difference between
  the phase of peak spectral color and of peak radiation flux, for
  pulsations generated by a hot spot on the surface of a neutron star
  spinning at 600~Hz, as a function of the stellar mass and
  radius. The remaining parameters of the calculation are the same as
  in Figure~\ref{fig:spectrum}. The color oscillations are introduced
  primarily by Doppler effects and, therefore, the contours on the
  left panel are nearly vertical, as in the right panel of
  Figure~\ref{fig:rms_600Hz}. Understanding the shape of the contours in
  the right panel of this figure is more subtle and is described in
  detail in the text.}
\label{fig:color_rms}
\end{figure*}

\begin{figure*}[t]
\centerline{
\psfig{file=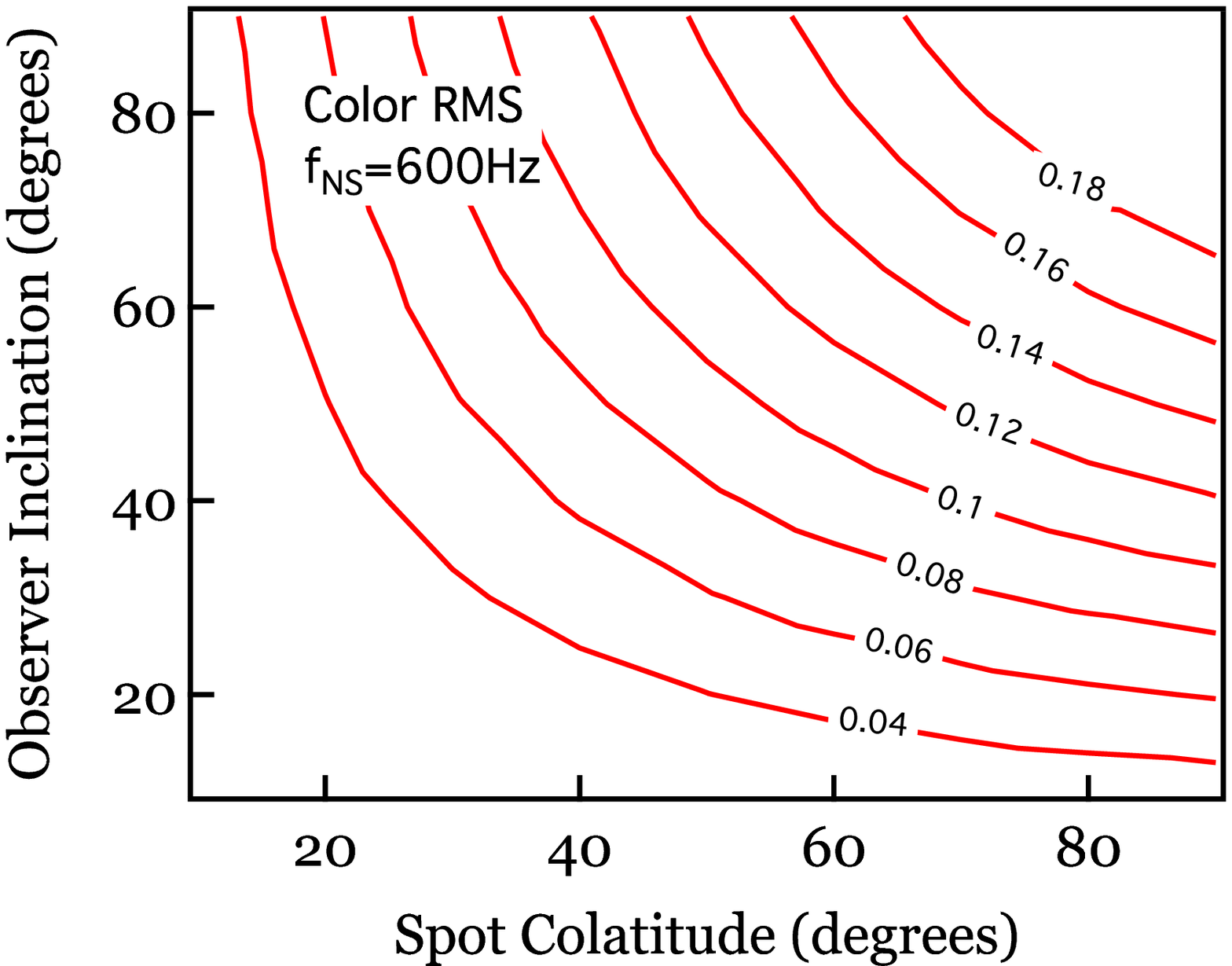,width=3.5in,clip=}
\psfig{file=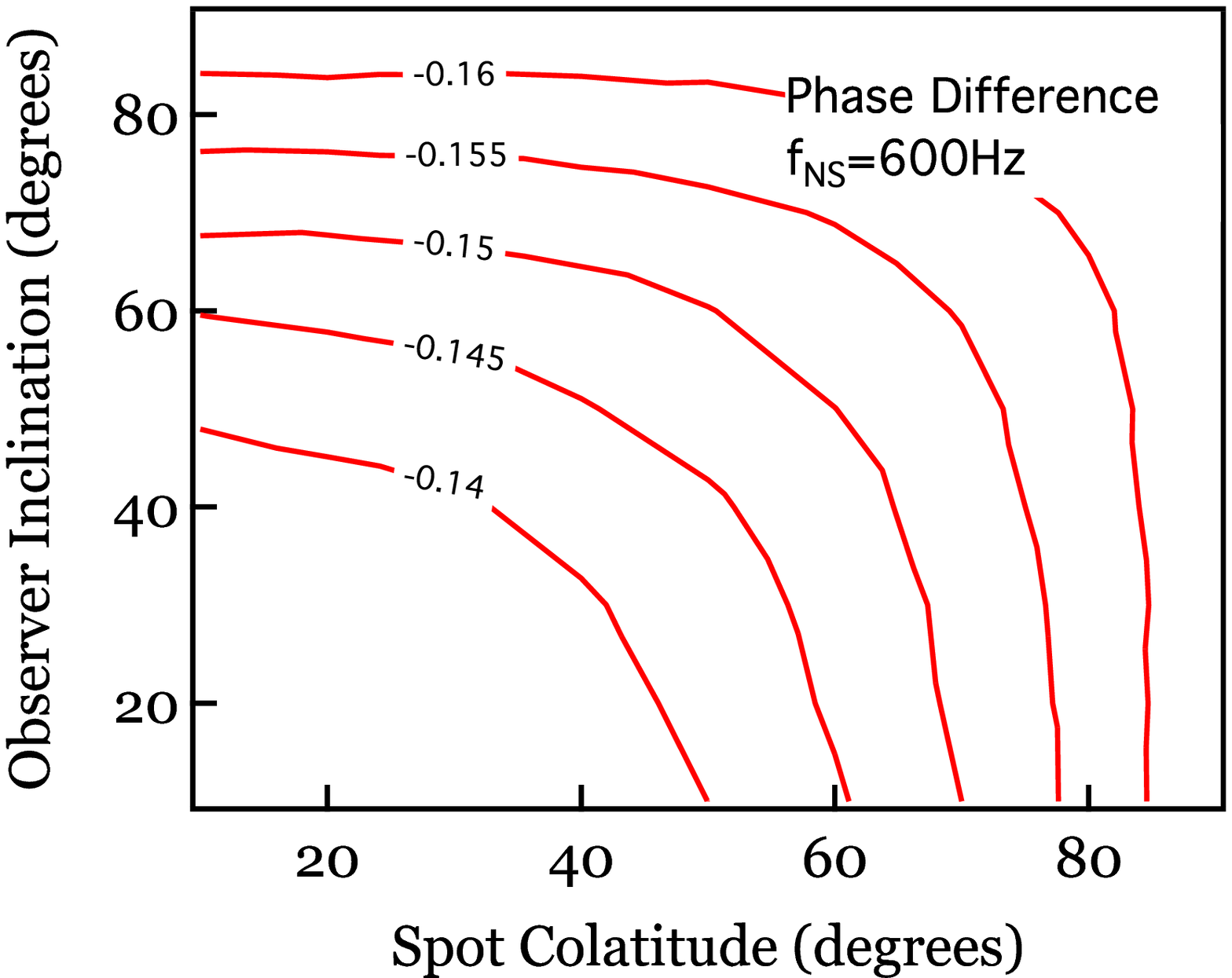,width=3.5in,clip=}}
\caption{Contours of constant {\em (Left)\/} fractional rms amplitude of a
  spectral color oscillation and {\em (Right)\/} difference between
  the phase of peak spectral color minus that of peak radiation flux,
  for pulsations generated by a hot spot on the surface of a neutron
  star spinning at 600~Hz, as a function of the colatitude of the spot
  and the inclination of the observer. The remaining parameters of the
  calculation are the same as in Figure~\ref{fig:spectrum}. The
  contours of the fractional rms amplitude of the spectral oscillation
  approximately follow curves on which the product $\sin i
  \sin\theta_{\rm s}$ is constant, as in the case of the first two
  observables shown in Figure~\ref{fig:rms_600Hz_theta}. In contrast,
  the phase difference results in a nearly orthogonal constraint and
  provides the information needed to measure the two angles
  independently.}
\label{fig:color_rms_theta}
\end{figure*}

\subsection{The Photon Energy Dependence of Pulse Profiles}

It is well understood that the Doppler effects that increase the
amplitudes of the higher harmonics in the pulse profile also introduce
a modulation to the radiation spectrum observed at infinity. This is
shown for a typical set of parameters in Figure~\ref{fig:spectrum} at
four different spin phases. As expected, the radiation spectrum is
softer in the range of spin phases for which the hot spot is receding
from the observer and harder when the hot spot is approaching.

We can quantify the degree of spectral modulation during a pulse phase
by defining a color as the ratio between the number of photons
observed in two energy bands separated at an energy equal to the
hot-spot temperature. In Figure~\ref{fig:color}, we plot the evolution
of such a color with pulse phase for a typical configuration, where we
use a photon energy equal to the hot-spot temperature (as measured at
the neutron-star surface) to separate the two energy bands. The
evolution of the color with pulse phase shows a nearly sinusoidal
modulation with a significant fractional rms amplitude and a peak
phase that is offset from that of the flux oscillation. As we will
show below, the amplitudes and the peak phases of the color
oscillations do not have the same dependence on the system parameters
as the amplitudes of the harmonics of the bolometric flux.  For this
reason, they provide the two additional observables needed to break
the degeneracies discussed in \S3.2 and allow a measurement of all the
system parameters. This is true even when using the minimal spectral
information encoded in one spectral color. If the number of photons
accumulated during an observation allow separating the pulse profiles
into several energy bands, additional consistency relations between
model parameters can be obtained.

Figure~\ref{fig:color_rms} shows contours of constant fractional
rms amplitude of the color oscillations (left panel) and of the
phase difference between the phase of peak color minus the phase of
peak flux (right panel) on the mass-radius parameter space.  If there
were no gravitational lensing and redshift effects, the amplitude of
the color oscillations would be strictly proportional to the radius of
the neutron star and would have a scaling similar to the one given in
equation~(\ref{eq:scaling}). However, gravitational effects introduce
a weak dependence on the neutron-star compactness such that the
contours of constant color amplitude shown in
Figure~\ref{fig:color_rms} are not parallel to the contours of
constant harmonic ratios shown in the right panel of
Figure~\ref{fig:rms_600Hz}.

The dependence of the contours of constant phase difference between
the color and flux oscillations on mass and radius shown in the right
panel of Figure~\ref{fig:color_rms} is more subtle. The peak phases of
the two oscillations are determined by the combination of the
evolution of the projected surface area of the hot spot on pulse phase
(which peaks at phase zero) and of the Doppler effects (which peak at
phase 0.75). The flux oscillation is determined predominantly by the
former effect, peaks close to zero phase, and the Doppler boosts
introduce small corrections, moving the peak toward earlier phases. On
the other hand, the color oscillation is determined predominantly by
the Doppler effects, peaks close to phase 0.75, and the surface area
projection introduces a small correction, moving the peak toward later
phases. For small values of the neutron-star compactness, for which
the gravitational lensing effects are weak, the relative shift between
the two peak phases is dominated by Doppler effects and, therefore,
the contours shown in Figure~\ref{fig:color_rms} become more
vertical. At high values of the neutron-star compactness, for which
the relative shift between the two peak phases is dominated by
gravitational lensing effects, the contours become nearly horizontal.

The contours of the four observables shown in Figures~\ref{fig:rms_600Hz} and 
\ref{fig:color_rms} on the mass-radius parameter space do not have the same 
dependence on the system properties. This leads to the conclusion
that measuring with sufficient accuracy the amplitudes of the lowest
two harmonics of the bolometric flux oscillation, as well as the
amplitude and relative phase of the spectral color oscillation is
adequate to uniquely determine all four parameters of each observed
system, as we will discuss in the next section.

We explore in Figure~\ref{fig:color_rms_theta} the dependence of the
fractional rms amplitude of the color oscillations (left panel) and
the phase difference between the flux and the color oscillations
(right panel) on the spot colatitude and the inclination of the
observer. The contours of constant fractional rms color amplitude lie
along curves on which the product $\sin i \sin\theta_{\rm s}$ is
nearly constant, as was the case with the two observables obtained
from the bolometric flux oscillations (see
Fig.~\ref{fig:rms_600Hz_theta}). This is expected given that the color
oscillations are also generated by Doppler effects and are determined
by the projection of the vector of the surface velocity along the
instantaneous line-of-sight between the observer and the hot spot. In
contrast, the phase difference between the flux and the color
oscillations shows a significantly different dependence on the two
angles. Therefore, if measuring the two angles independently is a goal
in and of itself, this last observable provides the additional piece
of information necessary to achieve it.

\begin{figure*}[t]
\centerline{
\psfig{file=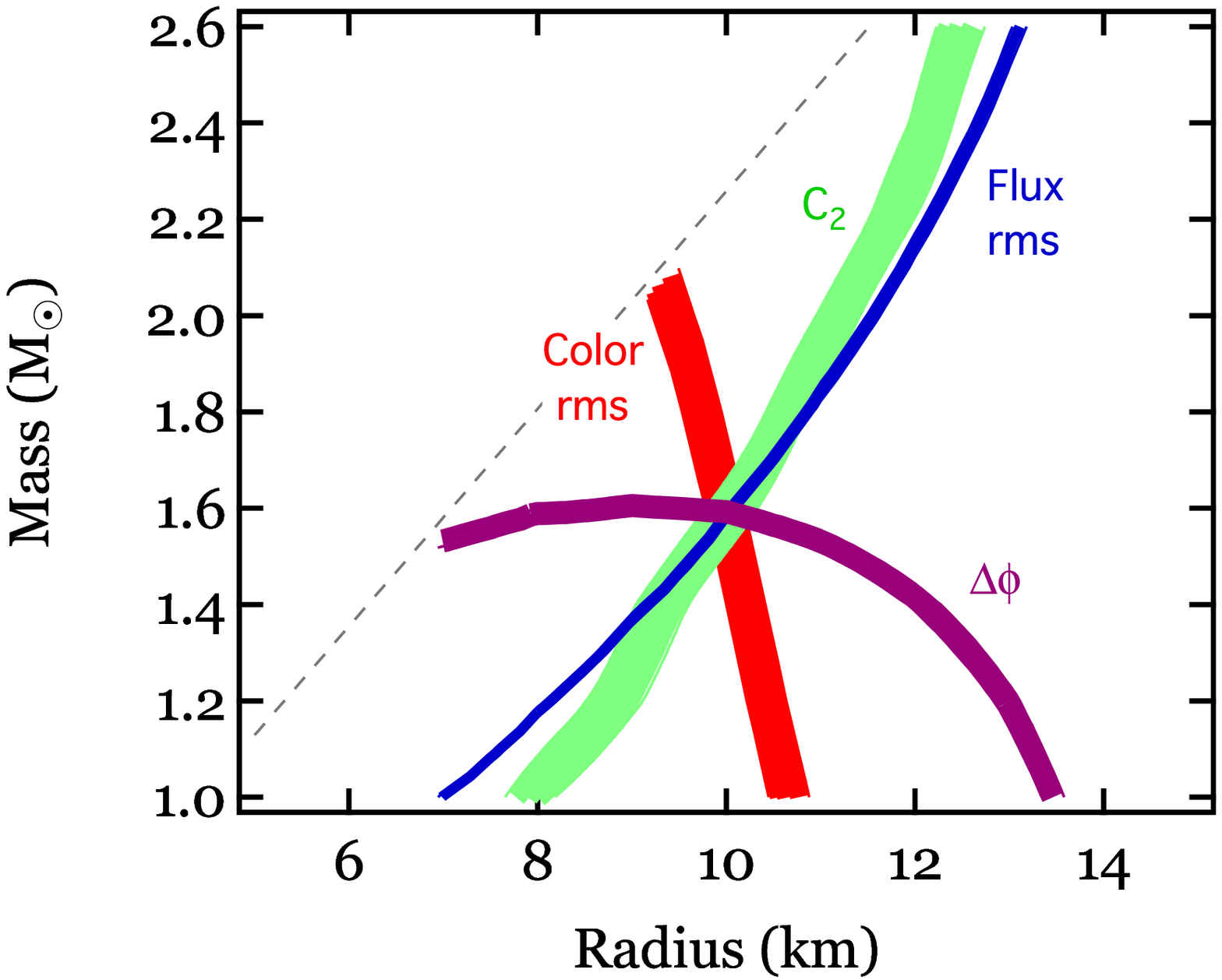,width=3.5in,clip=}
\psfig{file=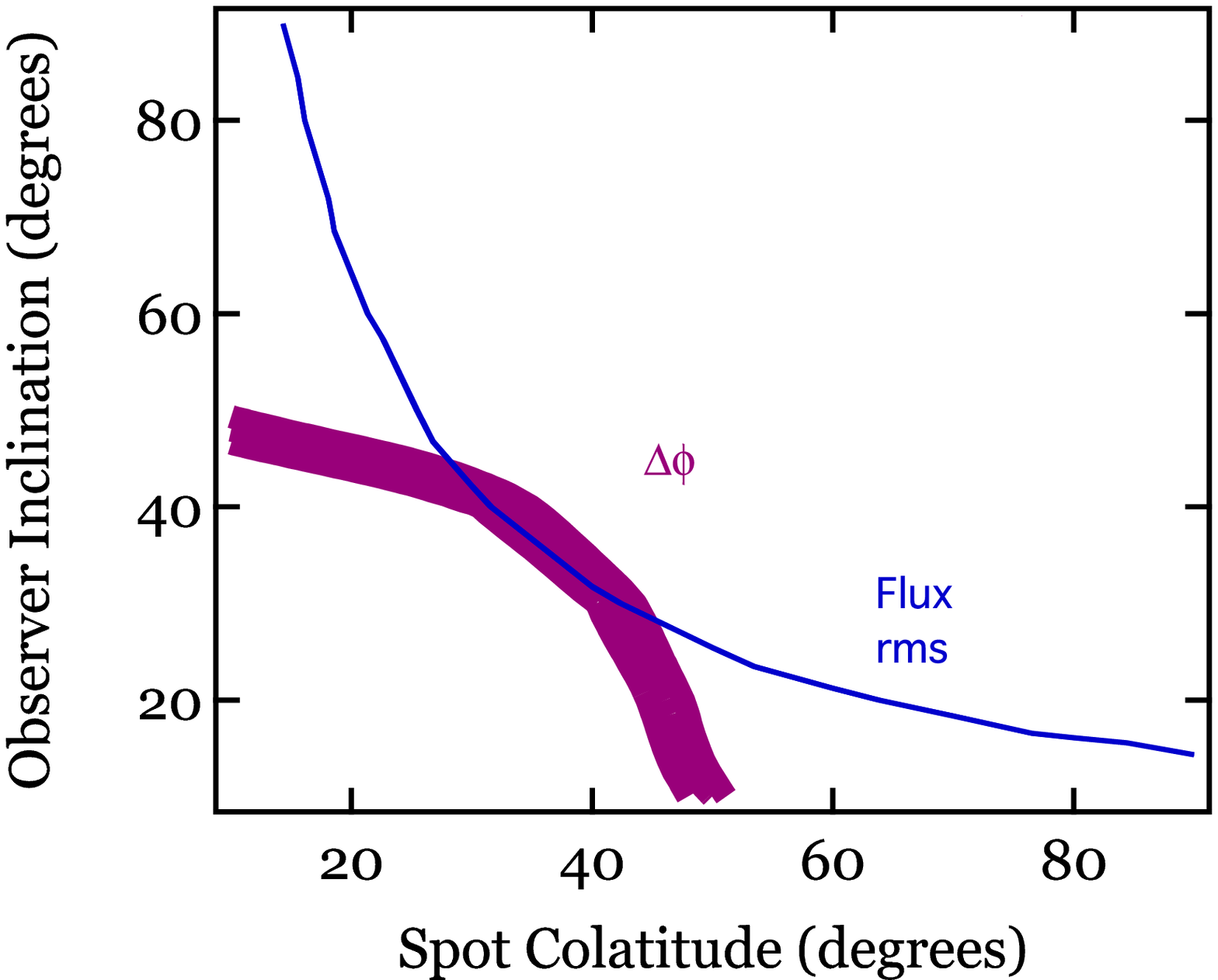,width=3.5in,clip=}}
\caption{Constraints on {\em (Left)\/} the neutron-star mass and
  equatorial radius and {\em (Right)\/} the inclination of the
  observer and colatitude of the hot spot, obtained from measuring
  four properties of a pulse profile. The measured values of the
  fractional rms amplitude of the pulse profile, the fractional rms
  amplitude $C_2$ 
  of the harmonic, the fractional rms amplitude of the color oscillation,
  and the phase difference $\Delta\phi$ between the flux and color
  oscillations were assumed to be equal to $0.17\pm 0.005$, $0.018\pm
  0.001$, $0.058\pm 0.001$, and $-0.136\pm 0.002$, respectively. The
  central values for each parameter correspond to the results of the
  calculation for a 1.6~$M_\odot$, 10~km neutron star, observed from
  an inclination of $30^\circ$, with a hot spot on its surface at a
  colatitude of $40^\circ$. Each panel represents a cut through the
  four-dimensional parameter space on which the parameters not shown
  remain constant. The distinct dependence of the four observables on
  the system parameters allows a unique recovery of the assumed
  neutron-star mass and radius.}
\label{fig:measurements}
\end{figure*}

\section{Prospects for Measuring Neutron-Star Properties from 
Pulse Profile Modeling}

In \S2, we showed that the pulse profile observed from a neutron star
spinning at a moderate rate can be accurately described by four
parameters: the mass of the neutron star, its equatorial radius, the
inclination of the observer, and the colatitude of the hot spot on the
stellar surface. In \S3, we demonstrated that observations of the
photon energy dependent pulse profiles result in at least four
measurable quantities that have a distinct dependence on the model
parameters: the amplitude of the bolometric flux oscillation, the
amplitude of its second harmonic, the amplitude of the spectral color
oscillation, and the phase difference between the bolometric flux and
the color oscillations.

The first three of these four observables depend on the same
combination $\sin i \sin\theta_s$ of the two geometric
parameters. Therefore, if the main goal of the pulse profile modeling
is to measure the masses and radii of neutron stars, this combination
can be treated as a single parameter. In this case, the first three
observables lead to a unique determination of the system
parameters. Nonetheless, if all four observables can be measured with
sufficient accuracy, then the geometry of the system can also be
uniquely determined.

In \S 3, we also quantitatively explored the dependence of the four
observables on the neutron star mass and radius. Three of the four
observables acquire detectable amplitudes because of the relativistic
Doppler shifts on the rapidly spinning neutron-star surface. As a
result, the accuracy of the measurements will depend on the accuracy
at which the amplitudes of the harmonics and color oscillations can be
measured. Using, for example, equation~(\ref{eq:scaling}), we can
relate an uncertainty $\Delta C_2$ for the measurement of the amplitude
of the harmonic to an uncertainty for the inferred radius $\Delta R_{\rm eq}$
as
\begin{equation}
\frac{\Delta R_{\rm eq}}{R_{\rm eq}}= 
\left[2C_1\left(\frac{2\pi f R_{\rm eq}}{c}\right)
\sin i \sin\theta_{\rm s}\right]^{-1}\Delta C_2 \;.
\label{eq:deltaR}
\end{equation}

As we show in the appendix, the uncertainty with which the amplitude
of a given harmonic can be measured depends on the total number of
source counts $S$ accumulated during the observations and on the total
number of background counts $B$ accumulated at the same time as
\begin{equation}
\Delta C_n=\frac{\sqrt{S+B}}{S}\;.
\label{eq:ampl_error}
\end{equation}
Combining the last two equations, we obtain
\begin{equation}
\frac{\Delta R_{\rm eq}}{R_{\rm eq}}=\left[
\left(\frac{4\pi f R_{\rm eq}}{c}\right)
\sin i \sin\theta_{\rm s}\right]^{-1} \left(\frac{\sqrt{S+B}}{C_1 S}\right)
\;.
\label{eq:final_unc}
\end{equation}
This equation provides the analytic understanding for the figure of
merit ${\cal R}$ introduced by Lo et al.\ (2013), which is closely
related to the quantity in the rightmost parentheses shown above.

When the number of source photons dominates that of the background,
i.e., when $S\gg B$, then the uncertainty in the measurement of the
neutron-star radius scales as
\begin{eqnarray}
\frac{\Delta R_{\rm eq}}{R_{\rm eq}}
&\simeq&0.055
\left(\frac{C_1}{0.3}\right)^{-1}
\left(\frac{f}{600~\mbox{Hz}}\right)^{-1}
\left(\frac{R_{\rm eq}}{10~\mbox{km}}\right)^{-1}\nonumber\\
&&\qquad
\left(\frac{\sin i}{0.5}\right)^{-1}
\left(\frac{\sin\theta_{\rm s}}{0.5}\right)^{-1}
\left(\frac{S}{10^6~\mbox{cts}}\right)^{-1/2}\;.
\end{eqnarray}
This relation suggests that achieving a 5\% accuracy in the
measurement of a neutron star radius from pulse profile modeling
requires accumulating of the order $10^6$ source counts.

Figure~\ref{fig:measurements} shows the constraints on the model
parameters that can be obtained from measuring the four observable
quantities discussed above with a precision that is characteristic of
an observation with one million source counts. In particular, we
assumed that all fractional amplitudes were measured with an accuracy
of $10^{-3}$ and the phase difference between flux and color
oscillations was measured with an accuracy of $5\times 10^{-4}$ (see
the Appendix).  For this example, the simulated lightcurve was generated
for a $1.6 M_\odot$, 10~km neutron star spinning at 600~Hz, observed
at an inclination of $30^\circ$, with a small uniform hot spot on its
surface at a colatitude of $40^\circ$.  The figure demonstrates that
the distinct dependence of the four observables on the system
parameters allows a unique recovery of the assumed neutron-star mass
and radius.  Moreover, it also shows that the assumed uncertainties
lead to a measurement of the neutron star mass and radius that is
sufficient to distinguish between different equations of state.

\subsection{Importance of the Background Model}

We emphasize that three of the four observables
discussed above are fractional rms amplitudes of Fourier
harmonics. Measuring these fractional amplitudes requires obtaining
both the pulsed and unpulsed (``DC'') components of the pulse profiles. If an
additional background that does not originate on the neutron-star
surface is present in the observed energy band, this component needs
to be separately measured and subtracted.  The alternative, i.e.,
measuring the properties of the additional background from the pulse
profiles themselves, introduces severe degeneracies between the
inferred model parameters, as shown in Lo et al.\ (2013). This is easy
to understand within the framework of counting system parameters and
observables that we followed here. Indeed, if we choose to perform
pulse profile analysis in two energy bands with unknown backgrounds,
then we introduce two additional parameters to our model, increasing
the total number to six. However, the number of observables that can
be inferred accurately from the pulse profiles remains equal to
four. This difference between the number of model parameters and
observables results in substantial degeneracies between the model
parameters of interest.

The approach we developed in this paper allows us to also investigate the
accuracy at which the number of background counts needs to be known
{\em a priori\/} in order for a desired accuracy in the mass and radius
measurement to be achieved. If we denote by $a_n$ the absolute amplitude
of the $n-$th Fourier component in a profile and by $N\equiv S+B$ the
total number of counts accumulated, then the fractional source amplitude
of the same Fourier component is simply
\begin{equation}
C_n=\frac{a_n}{S}=\frac{a_n}{N-B}\;.
\end{equation}
At least two effects, in principle, contribute to the uncertainty in
the measurement of the fractional source amplitude: the uncertainty in the
measurement of the absolute amplitude $\Delta a$ and the uncertainty
in the {\em a priori\/} knowledge of the background $\Delta B$.
Incorporating both sources of error, we obtain
\begin{eqnarray}
\Delta C_n^2&=&\frac{\Delta a_n^2}{(N-B)^2}+
   \left[\frac{a_n}{(N-B)^2}\right]^2\Delta B^2\nonumber\\
&=&\left(\frac{\Delta a_n}{S}\right)^2+
   \left(\frac{a_n}{S^2}\right)^2\Delta B^2\;.
\end{eqnarray}
The first term in the right-hand side of this equation is the Poisson
error in the measurement of the fractional amplitude and is given by
equation~(\ref{eq:ampl_error}). Assuming that the uncertainty in the
measurement of the background is Poisson dominated, $\Delta B=\sqrt{B}$.
Inserting these two expressions in the last equation, we obtain
\begin{equation}
\Delta C_n^2=\left(\frac{\sqrt{S+B}}{S}\right)^2+
   C_n^2\frac{B}{S^2}
\end{equation}
or simply
\begin{equation}
\Delta C_n=\frac{\sqrt{S+B(1+C_n^2)}}{S}\;.
\label{eq:back_ampl}
\end{equation}

Equation~(\ref{eq:back_ampl}) shows that the background contributes in
two ways in the uncertainty of the measured fractional amplitude of
the source: the overall counts in the background increase the level of
the Poisson noise in the power spectrum and hence degrade the
measurement of the absolute amplitude of the pulsations. At the same
time, the uncertainty in the subtraction of the background counts
affects the inference of the fractional amplitude of the pulsations.
Because $C_n^2\ll 1$, the latter effect is always subdominant compared
to the former. We can, therefore, neglect it and simply use
equation~(\ref{eq:final_unc}) to infer the expected uncertainty in the
measurement of neutron-star radii when the observations have a
significant background.

\section{Conclusions}

In this paper, we investigated how pulse profiles generated by hot
spots on moderately spinning neutron stars can be used to infer the
stellar mass and radius. We showed that bolometric pulse profiles do
not contain sufficient information to break parameter degeneracies to
uniquely measure these quantities. However, a measurement of the
spectral color oscillations provides additional constraints that allow
a separate determination of $M$ and $R_{\rm eq}$. Extracting pulse
profiles even in just two different energy bands is sufficient to
derive this information.  However, achieving $5\%$ precision in
neutron-star radius requires accumulating $\gtrsim 10^6$~counts in the
pulse profile measurements. This can be accomplished by long exposure
times (as in the case of {\em NICER}) or by a large collecting area
(as in the case of {\em LOFT}).

The requirements we presented here for making measurements of the
neutron star radius with a given precision are robust to the detailed
energy coverage and response of a particular instrument. Given that
our study has utilized idealized light curves, the details of, e.g.,
required number of counts, can be refined for a particular detector or
choice of energy bandpass. However, we note that because only two
energy channels are required to make the spectral color oscillation
measurement, even a modest energy resolution is sufficient. On the
other hand, a broad energy coverage is advantageous because the color
oscillations are more pronounced when measured over a wider energy
range.  Ideally, the energy bandpass should include at least the
blackbody peak or the exponential tail above the peak.

X-ray spectral color oscillations have been previously reported in
several thermonuclear burst oscillations from two neutron stars
(Strohmayer et al. 1999; Strohmayer 2000).  These color oscillations 
were all measured during the burst decay phase and were found to be in 
phase with the flux oscillations, contrary to the expectations discussed 
above. This suggests that the color oscillations in the burst tails are not 
generated by Doppler effects and are not appropriately modeled by the hot spot 
model described above. This also corroborates other lines of arguments that 
the oscillations in the burst tail are generated by a different mechanism 
(e.g., surface modes) than those in the burst rise (see also Watts 2012 and 
references therein). 

In the case of oscillations observed during the rise phases of X-ray 
bursts, which are the prime targets for {\em LOFT\/}, the predominant 
background arises from the X-ray emission from the accretion flow. In 
the case of surface emission from rotation powered pulsars, which are 
the main targets for {\em NICER\/}, the non-thermal emission from the
neutron-star magnetosphere is the main source of the background. In
both cases, the energy spectrum of the pulsations is very different
from the energy spectrum of the background. If the shape of the energy
spectrum of the background is known, e.g., from theoretical models and
prior observations, then its overall normalization can be measured at
an energy band where the pulsed surface emission is negligible, i.e.,
at hard X-rays. Such an approach will lead to an independent
measurement of the background in the energy bands of interest and will
not adversely affect measuring neutron-star masses and radii from
pulse profile modeling.

\acknowledgements We thank the {\em NICER\/} and {\em LOFT\/} science
teams for many discussions on ray tracing in neutron-star spacetimes.
This work was supported in part by NSF grant AST-1108753, NSF CAREER
award AST-0746549, and {\em Chandra\/} Theory grant
TM2-13002X. F.\"O. gratefully acknowledges support from the Radcliffe
Institute for Advanced Study at Harvard University.  D.P.\ and
F.\"O.\ thank the Institute for Theory and Computation at Harvard
University for their hospitality during the time that this work was
completed.

\appendix

\section{Uncertainties in the measurement of pulsation amplitudes and phases}

In this appendix we show, for completeness, that the accuracy with
which the fractional amplitude and phase of a periodic signal can be
measured from a noisy time series that is folded at the known period
of the signal depends on the total number of counts due to the source
and due to the known background (see also discussion in van der Klis
1989).

\begin{figure}[t]
\centerline{\psfig{file=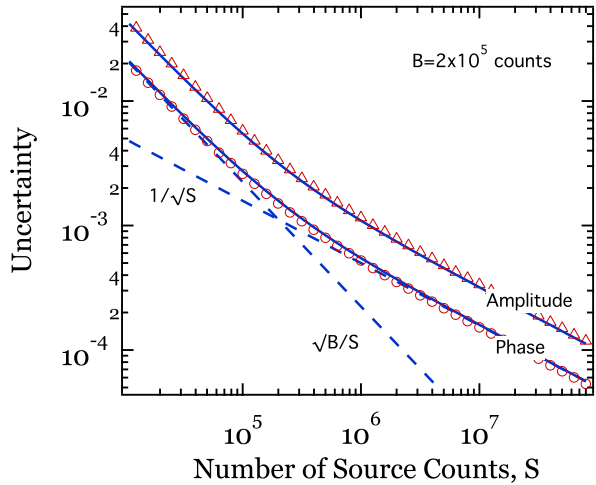,width=3.5in,clip=}}
\caption{The dependence of the uncertainty in the measured amplitude
  {\em (triangles)\/} and phase {\em (circles)\/} of a periodic signal
  on the number of photons, $S$, detected from a source. For the
  simulations shown in this Figure, the number of background photons
  was set to $B=2\times 10^5$. The solid lines show the two analytic
  expression for the uncertainties, as given in the text. When the
  observation is background-dominated, the uncertainty scales as
  $\sqrt{B}/S$ whereas, when it is source-dominated, it scales as
  $1/\sqrt{S}$.}
\label{fig:amplitudes}
\end{figure}

Because of Poisson noise in the measurement, the amplitudes of the
Fourier components of a signal that is assumed to be constant in time
are not zero. If we denote by $x_k$ the number of photons measured in
the $k$-th bin of such a time series and by $a_{j}$ the Fourier
component at frequency bin $j$, where $n_{\rm b}$ is the number
of bins, we can write
\begin{equation}
x_k=\frac{1}{n_{\rm b}}
\sum_{j=-n_{\rm b}/2}^{n_{\rm b}/2}a_j\exp\left(-2\pi ijk/n_{\rm b} \right)\;.
\end{equation}
The mean number of photons in each bin is
\begin{equation}
\bar{x}_k=\frac{a_0}{n_{\rm b}}=\frac{S+B}{n_{\rm b}}\;,
\end{equation}
where we have denoted by $S$ and $B$ the total number of photons collected
due to the source and due to the background, respectively.

Poisson noise leads to a flat Fourier spectrum with a mean amplitude
that we denote here by $a_{\rm P}$. We can calculate this mean
amplitude using Parseval's theorem. The variance in the number of
counts collected in each time bin is related to the Fourier amplitudes
by
\begin{equation}
\mbox{Var}(x_k)=\frac{1}{n_{\rm b}} \sum_{j=-n_{\rm b}/2;j\ne 0}^{n_{\rm b}-1}
\vert a_j\vert^2\simeq \vert a_{\rm P} \vert^2;. 
\end{equation}
Moreover, the standard deviation $\sigma$ of the number counts in each bin
is related to the mean number of photons by $\sigma^2=\bar{x}_k$ and to
the variance by
\begin{equation}
\sigma^2=\frac{\mbox{Var}(x_k)}{n_{\rm b}}\;.
\end{equation}
Combining the last two equations together, we find that the mean
amplitude of the Fourier components of the Poisson noise is
\begin{equation}
\vert a_{\rm P}\vert=\frac{\sqrt{S+B}}{n_{\rm b}}\;.
\end{equation}
If the time series contains in addition to the Poisson noise a
periodic signal that is uncorrelated with the noise, the quantity
$\vert a_{\rm P}\vert$ will represent the uncertainty within which the
amplitude of the signal can be measured.

In this paper, we have been using fractional amplitudes of the
periodic signals, which we can obtain by dividing the absolute
amplitude of a signal by the average number of counts in each bin due
to the source alone. In this case, the noise level $C_{\rm P}$ of the
fractional amplitudes becomes
\begin{equation}
C_{\rm P}=\left(\frac{\vert a_{\rm P}\vert}{n_{\rm b}}\right)
\left(\frac{S}{n_{\rm b}}\right)^{-1}=\frac{\sqrt{S+B}}{S}\;.
\label{eq:error_ampl}
\end{equation}

In order to verify the validity of equation~(\ref{eq:error_ampl}), we
generated a large number of simulated observations using the
lightcurve shown in Figure~\ref{fig:color} for a different number of
source photons $S$ and a constant background of $B=2\times
10^5$~counts, adding the appropriate level of Poisson noise. For each
simulated observation, we measured the amplitude of the fundamental
and of the second harmonic and used the distribution of amplitudes over
the various realizations to infer the uncertainty in each measurement.
The open triangles in Figure~\ref{fig:amplitudes} show the measured
uncertainty of the amplitudes as a function of the number of source
counts. The solid line that follows the open triangles is the analytic
result of equation~(\ref{eq:error_ampl}) and agrees well with the 
results of the simulations.

We used the same simulations to also measure the phase of the
fundamental Fourier component with respect to a fiducial phase in all
realizations of the data. The open circles in
Figure~\ref{fig:amplitudes} show the dependence of the uncertainty in
the measured phase on the number of source counts. The solid line that
follows the open circles is given by the simple relation
\begin{equation}
\sigma_\phi=\frac{\sqrt{S+B}}{2S}
\end{equation}
and also agrees well with the results of the simulations.

\end{document}